\documentclass[12pt]{iopart}
\usepackage{graphicx}
\usepackage{epsf}
\usepackage{amsmath}
\usepackage{amsfonts}
\usepackage{amssymb}
\usepackage{epstopdf}
  \usepackage{hyperref}
\usepackage{color}
\usepackage{verbatim}
\usepackage{multirow}
\usepackage{bm}
\usepackage{orcidlink}

\begin{document}
\title[Solving TOV equations in $f(T)$ gravity]{Solving Tolman-Oppenheimer-Volkoff equations in $f(T)$ gravity: a novel approach}

\author{H G M Fortes\orcidlink{0000-0002-2069-2628}$^1$ and J C N Araujo\orcidlink{0000-0003-4418-4289}$^1$}

\address{$^1$ Divisão de Astrofísica, Instituto Nacional de Pesquisas Espaciais, Avenida dos Astronautas 1758, 12227-010 SP, Brazil}

\ead{hemily.gomes@gmail.com}

\begin{abstract}
The torsion models have stood out among the proposals for an alternative description of gravity. The simplest of them, the Teleparallel theory, is equivalent to General Relativity and there are many studies dealing with its extension to more general functions of the torsion $T$. The purpose of {our study } is to consider a family of $f(T)$ models and apply their corresponding Tolman-Oppenheimer-{Volkoff} equations to compact objects such as neutron stars. {Thus, through a numerical analysis, calculate, among other things, the maximum mass allowed by the model for a neutron star, which also allows us to evaluate which models agree with the observations.} In the present paper, the first in the series, we show explicitly the set of equations that must be solved, and how to solve it, in order to model compact stars in $f(T)$ gravity without the need to adopt any particular form for the metric functions or consider any perturbative approach, as has been done in some works in the literature. {Examples are given of how our approach works, modelling polytropic stars. We also show that some numerical instabilities reported in a previous study by other authors do not appear in our novel approach. This is an important advance, since it is possible to answer an issue not responded in a previous study, because numerical instabilities {prevented} proceeding with the calculations. Last but not least, we explicitly show the torsion behavior inside and outside the star. This is an important question, because with this study we can understand the role of torsion in the structure of the star.}
\end{abstract}


\submitto{\CQG}
\maketitle

\section{Introduction}
\label{int}

The theory of General Relativity (GR) with a cosmological constant has offered an explanation for the accelerated expansion of the universe {in the context of the $\Lambda$ CDM model}, where it is also assumed the existence of the so-called dark matter whose properties are still unknown. In addition to the lack of knowledge about dark matter, there is also the problem of the cosmological constant whose value is very small compared to the predictions of quantum field theory.

In an attempt to circumvent the theoretical impasses and offer an alternative explanation for the observed acceleration, it is necessary to make changes in the fundamental theory. In this sense, there have been several proposals for alternative theories of gravitation based on modifications of the GR, for example, adding to the Lagrangian terms of mass or non-linear terms in curvature, among others.

The so-called Teleparallel theory of Gravity \cite{Aldro}-\cite{TT} is known to be equivalent to GR, with the basic difference of having in its formulation the torsion scalar $T$ instead of the curvature scalar $R$. {It is worth mentioning that the equivalence is complete. The conservation law of the source energy-momentum tensor and also the equations of motion in the presence of matter fields are equivalent in both theories.}


Thus, we can consider changes in both GR and Teleparallel theory in order to build alternative models of gravitation. An important family of modifications of the Einstein-Hilbert action are the theories $f(R)$ where we have a function of the scalar of curvature as Lagrangian density. However, the field equations of these models are {in general} fourth order differential equations, which makes their analysis difficult. It is possible, then, to make equivalent modifications in the Teleparallel theory considering the Lagrangian density as a function of the torsion scalar. These models, in turn, have the advantage of taking us to second order field equations, therefore, simpler than those obtained in $f(R)$. It is worth mentioning that the theory $f(T)$ has presented interesting cosmological and astrophysical solutions, providing alternative interpretations for the acceleration phases of the universe \cite {Ferraro}-\cite{Karami}. 

{Generalized Tolman-Oppenheimer-Volkoff} (TOV) \cite{TOV} equations are derived from the theory of gravity and can be used to model the structure of a spherically symmetrical object that is in hydrostatic equilibrium. From the Einstein-Hilbert action of GR, {equations are} obtained and can be used to describe, for example, neutron stars. With this, it is possible to determine, among other things, the maximum mass allowed for this type of star {for a given equation of state}. In this sense, considering different gravity models could, in principle, lead us to different maximum masses, see \cite{Nunes}. This would be important, for example, to explain one of the components involved in the event GW190814 \cite{GW} concerning the coalescence involving a black hole of 22.2–24.3 $M_\odot$ and a compact object of mass 2.50–2.67 $M_\odot$, the latter being a relatively high-mass neutron star or a low-mass black hole. Therefore, it is evident the great relevance of considering alternative models of gravitation, such as the $f(T)$ models in the description of compact objects.

The goal of our study is to propose a specific family of $f(T)$ models and consider their applications in some astrophysical contexts, namely, the description of compact objects such as neutron stars. Such models must present solutions for these objects with a mass-radius relation that is in agreement with the observations. For this, it is necessary to obtain the TOV equations corresponding to the model in question and look for solutions of these equations. The proposal is to obtain these solutions for neutron stars from a more general approach than those already presented in the literature \cite{Ganiou}-\cite{Pace}.

The TOV equations for $f(T)$ models will depend on the form assumed for this function and, together with the equation of state of neutron star, lead us to a system of coupled differential equations of non-trivial resolution. Thus, many authors choose to consider perturbative cases, approximations, assume conditions on the parameters of the theory or restrict themselves to very simple forms for $f(T)$, which is avoided with the approach presented in this paper. {It is worth mentioning that in \cite{Ilijic}, the authors seem to be among the few who solve the problem appropriately for a particular $f(T)$.} 

{In fact, one will see later on that the present paper improves Ref. \cite{Ilijic} in some important aspects. First, in an issue related to numerical instabilities. Second, in an issue regarding the maximum mass, which is closely related to the non-appearance of certain instabilities in our numerical calculations that appear in the calculations by \cite{Ilijic}. Third, we explicitly show the behaviour of the torsion scalar inside and outside the star.}

In the present paper, the first in the series, we explicitly show the set of equations that must be solved, and how to solve it, in order to model compact stars in $f(T)$ gravity without the need to adopt any particular form for the metric functions or consider any perturbative approach. This is an essential procedure which allows the readers to extend more easily the approach to other $f(T)$. More specifically, Section \ref{tetrada} considers the basic equations; in Section \ref{sec 3} we present the equations to be solved to model compact stars in f(T) gravity; in Section \ref{sec 4} numerical examples are presented and the final remarks are considered in Section \ref{sec 5}.

\section{The basic equations of f(T) gravity for spherically symmetric metric}
\label{tetrada}

Theories with torsion are formulated from an alternative dynamical variable ${h^a}_\mu$, the tetrad, which is a set of four vectors that define {a local inertial frame} at each point  \cite{TT}. The metric $g_{\mu \nu}$, usually adopted in models with curvature, and the tetrad ${h^a}_ \mu$ are related as follows:
\begin{equation}
g_{\mu\nu}={h^a}_\mu \, {h^b}_\nu\,  \eta_{ab} \ ,
\end{equation}
where $\eta_{ab} = \mbox{diag}(+ 1, -1, -1, -1)$ is the Minkowski metric of the tangent space.

Gravity theories constructed from the metric will always be invariant under local Lorentz transformations. On the other hand, torsion models will not necessarily be, which will generally depend on a choice of tetrads and appropriate spin connection.
In addition, a non-invariant theory will be sensitive to the choice of the tetrad and may lead to different solutions. Thus, this choice is fundamental for the good behavior of the $f(T)$ models, which has been extensively studied in recent years \cite{TT,Tamanini} and is not always taken into account when working with models with torsion.

{This was so till the seminal paper \cite{Krssak}, in which a covariant $f(T)$ gravity is formulated {(see also \cite{GKS})}. 
Thus, the problems related to the violation of local Lorentz invariance and frame dependence 
could be resolved. To obtain a covariant $f(T)$ gravity they also considered the inertial spin connection as a dynamical variable. Recall that in the usual formulation (hereafter referred to original formulation) only the tetrad {can be} considered as dynamical variable.}

{It is also worth mentioning that in the covariant $f(T)$ formulation by \cite{Krssak} one is allowed to adopt an arbitrary tetrad and then obtain its corresponding spin connection. {Thus, there is no need to worry about bad or good tetrads\footnote{Following Ref \cite{Tamanini}, good tetrads are those that do not impose any restrictions in the form of $f(T)$. If any restriction is imposed in the form of $f(T)$, you have a bad tetrad.}, as they all lead to the same field equations.} An interesting issue is how to obtain the spin connection, but since this is out of the scope of the present article, we refer the reader to \cite{Krssak} for detail.} {It is noteworthy that although  the covariant formulation is mentioned here, it is not used in any derivation in the present paper.}

{Before proceeding, we refer the reader to the interesting recent paper by \cite{Golovnev}, who discusses issues regarding the Lorentz-invariance in $f(T)$ gravity and its covariant formulation.}
 
For the purpose of this article, we consider the $f(T)$ models in spherically symmetric spacetime \cite{SSS1}-\cite{SSS12} derived from the original formulation.  A suitable tetrad in this formulation reads \cite{Tamanini}:

\begin{multline}
{h_\mu}^a=
\left(
\begin{array}{cc}
 e^{A/2} & 0 \\
 0 & e^{B/2} \sin \theta \cos \phi  \\
 0 & -r \left(\cos \theta \cos \phi \sin \gamma+\sin \phi \cos\gamma \right) \\ 
 0 & r \sin \theta \left(\sin \phi \sin\gamma -\cos \theta \cos \phi \cos\gamma \right)
\end{array}\right.\\
\left.
\begin{array}{cc}
 0 & 0 \\
 e^{B/2} \sin \theta \sin \phi  & e^{B/2} \cos \theta  \\
r \left(\cos \phi \cos\gamma -\cos \theta \sin \phi \sin\gamma \right) & r
   \sin \theta \sin\gamma  \\
 -r \sin \theta \left(\cos \theta \sin \phi \cos\gamma +\cos
   \phi \sin\gamma \right) & r \sin ^2\theta \cos\gamma 
\end{array}
\right) \, ,
\label{006}
\end{multline}
where $A$, $B$, $\theta$, $\phi$ are those from the spherically symmetric metric $ds^2=e^{A(t,r)}\, dt^2-e^{B(t,r)}\, dr^2-r^2\, d\theta^2-r^2 \sin ^2 \theta \, d\phi^2$, and $\gamma=\gamma(r)$ is one of the Euler's angles. {The tetrad above is obtained from a general three-dimensional rotation parametrized by its Euler angles. In general, even if they are taken to be arbitrary
functions of the spherical coordinates $t, r, \theta, \phi$, the transformed tetrad returns the spherically symmetric metric. Thus, for simplicity, we choose $\gamma=\gamma(r)$. See \cite{Tamanini} for more details.}

{It is worth noting} that Ref. \cite{Tamanini} shows how employing nondiagonal (rotated) tetrads permits us to recover spherically symmetric solutions in vacuum and to prove Birkhoff’s theorem, using the same field equations derived
in our paper. It is well known that different tetrads lead to different physical theories in $f(T)$ gravity, so this result holds specifically for tetrad (\ref{006}). However, it contains both
the tetrads considered in \cite{Kpa,Pace}, for example. See, also, \cite{SSS1}-\cite{SSS9}.

Since \cite{Tamanini} was one of the first papers to obtain the correct field equations for spherically symmetrical spacetimes, we will adopt it to obtain the basic equations used here to model compact stars in $f(T)$ gravity.

The torsion tensor is built from the antissymmetry part of the connection {only,}
\begin{equation}
  T^{\sigma}{}_{\mu\nu} = \Gamma^{\sigma}{}_{\nu\mu} - \Gamma^{\sigma}{}_{\mu\nu} =
  h_i{}^{\sigma} (\partial_\mu h^i{}_{\nu}-\partial_\nu h^i{}_{\mu}) \, ,
\end{equation} 
{since the spin connection is null in the original formulation; the contorsion is defined by}
\begin{equation}
    {K^{\mu\nu}}_{\sigma} = - \frac{1}{2} \left({T^{\mu\nu}}_{\sigma} - {T^{\nu\mu}}_{\sigma} - {T_{\sigma}}^{\mu\nu}\right).
\end{equation}
A useful tensor reads
\begin{equation}
    {S_{\sigma}}^{\mu\nu} = \frac{1}{2} \left({K^{\mu\nu}}_{\sigma} + \delta^\mu_\sigma {T^{\rho\nu}}_{\rho}- \delta^\nu_\sigma {T^{\rho\mu}}_{\rho} \right).
\end{equation}
Finally, the torsion scalar T is defined by (see, e.g., \cite{Bohmer})
\begin{equation}
    T =  {S_{\sigma}}^{\mu\nu} T^{\sigma}{}_{\mu\nu}.
\end{equation}

By using the tetrad (\ref{006}), we can write $T(r)$ as:
\begin{equation}
  T(r) = \frac{2\, e^{-B}}{r^2} 
  \Bigl[ 
    1 + e^B + 2\,e^{B/2} \sin\gamma + 
    2\, e^{B/2}\, r\,  \gamma' \cos\gamma +
    r\, A' \left(1+e^{B/2}\sin\gamma\right) 
  \Bigr]\,,
  \label{008}
\end{equation}
where the prime stands for derivative with respect to $r$.

By varying the action for the $f(T)$ model 

{
\begin{equation}
S = \int   \ \left( \frac{f(T)}{16\pi} + \mathcal{L}_m \right) \ {\rm det}({h_\mu}^a)\ d^4 x \ ,
\end{equation}
}
{with respect to the Vierbein components $h_\mu^a(r)$, {where $\mathcal{L}_m$ is the Lagrangian density of matter fields,} one can obtain the following field equations \cite{Tamanini}:}
\begin{eqnarray} \hspace{-1.5 cm}
  4\pi P = -\frac{f}{4} +\frac{f_T\,e^{-B}}{4r^2} \left( 2-2\,e^B+r^2e^B T-2r\,A' \right) 
\label{026}\\ \hspace{-1.5 cm}
4\pi\rho = \frac{f}{4} -\frac{f_T\,e^{-B}}{4r^2}\left( 2-2\,e^B+r^2e^B T-2r\,B' \right) 
  -\frac{f_{TT}\,T'e^{-B}}{r} \left(1+e^{B/2}\sin\gamma\right) 
\label{025}\\ \hspace{-1.5 cm}
  f_{TT}\,T'\cos\gamma = 0 \label{007}\\ \hspace{-1.5 cm}
  f_T\,\dot B = 0  \label{009}\\ \hspace{-1.5 cm}
  \dot{B} \Big[ 2\,f_{TT}\left(1+e^{B/2}\sin\gamma\right) \left(2-2\,e^B+r^2e^B T+ 2r\, A'\right)
 + e^B r^2 f_T\Big] +\nonumber \\ \hspace{-1.5 cm} - 4r\,f_{TT}\,\dot{A}'\left( 1+e^{B/2}\sin\gamma\right)^2 =0 
\label{024}\\ \hspace{-1.5 cm}
  f_{TT}\Bigl[ -4e^A r T'(1+e^{B/2} \sin{\gamma})-\dot{B}^2\left( 2-2e^B+r^2e^B T \right) 
  -2rA' \left(e^ArT'+\dot B^2\right) +\nonumber \\ \hspace{-1.5 cm} +4r \dot B\dot A' \left(1+e^{B/2}\sin\gamma\right) \Bigr] 
  +f_T \Big[ 4\,e^A-4e^Ae^B-e^Ar^2A'^2+ 2\,e^Ar\,B' + \nonumber \\ \hspace{-1.5 cm}
  +e^Ar\,A' \left( 2+r\,B' \right) -2r^2e^A A''
  - e^Br^2 \dot A\dot B 
  +e^Br^2 \dot B^2 +2e^Br^2\ddot B \Big] =0
\label{027}
\end{eqnarray}
where $f_T=\frac{\partial f}{\partial T}$ and $f_{TT}=\frac{\partial f_T}{\partial T}$ and the overdots stand for differentiation with respect to $t$. {Note that the above action variation regarding matter fields leads to the usual energy-momentum tensor.}{ The source term is given by an isotropic perfect fluid, with pressure P and energy density $\rho$.}

{Additionally, the well-known conservation equation, which also holds in $f(T)$ gravity (see \cite{Bohmer} for a detailed derivation), reads:
\begin{equation}
    A' = -2 \frac{P'}{P+\rho},
\label{ce1}    
\end{equation}
which will be very useful for our purposes later on.}

\section{Compact stars on $f(T)$ models}\label{sec 3}

Since we are interested to model compact stars in hydrostatic equilibrium in $f(T)$ gravity, $A$ and $B$, and consequently all other quantities such as $P$ and $\rho$, do not depend on time. As a result, equations (\ref{009}) and (\ref{024}) are identically null and several terms in equation (\ref{027}) are taken equal to zero. The resulting equations are complicated anyway and it does not seem possible to get a set of equations similar in structure to the TOV {equation that comes} from General Relativity. 

In fact, there are papers in the literature related to compact stars in $f(T)$ gravity \cite{Ganiou,Kpa,Pace,Pace2} in which very complicated equations are obtained for different tetrads, as well as for some {\it ansatz} for $B$ and or $T$, for example.

The main objective here is, then, to obtain the system of differential equations adequate for this type of analysis and, without assuming restrictions for the functions $A(r)$, $B(r)$, etc., find their solutions without approximations or perturbative calculation. 

In principle, we are going to consider, as an example, $f(T)$ given by the following form
\begin{eqnarray}
f(T)=T + \xi \, T^2 \, ,
\label{fT}
\end{eqnarray}
where $\xi$ is an arbitrary real. However, the procedure could be repeated later on for even more general forms of $f(T)$. Notice that, for $\xi=0$, the results from the Teleparallel equivalent  of  General  Relativity can be retrieved.

{ This specific choice is the simplest one and it is inspired in the Starobinsky model in $f(R)$ gravity, which has this same functional form. However, following the approach presented in this paper, it is straightforward to consider other functional forms for $f(T)$.}

{Additionally, taking $\gamma=-\pi/2$\footnote{The specific choice $\gamma=-\pi/2$ will allow us to reproduce the TOV GR equation in the $\xi=0$ limit.} in order to get rid of the restriction on $f(T)$ from (\ref{007}) and by considering the hydrostatic equilibrium ($\dot{B}=0$, $\dot{A}=0$), equations (\ref{025})-(\ref{027}) can be rewritten as:}
\begin{eqnarray} \hspace{-2 cm}
\frac{(1-e^B)}{2}-4\pi\, P \, r^2\,  e^{B}+\frac{\xi}{r^2}\left(6+3e^{-B}-e^B-8e^{-B/2}\right)+ \nonumber \\ \hspace{-2 cm}
+\frac{1}{2\, r}\biggl[r^2+12\xi\left(1+e^{-B}
-2e^{-B/2}\right)\biggr]A' 
+\xi\left(3e^{-B}-4e^{B/2}+1\right) {A'}^2
=0 \label{eq1}
\end{eqnarray}

\begin{eqnarray} \hspace{-2 cm}
-\frac{\left(1-e^B\right)
}{2r^2}-4\pi\rho e^B+\frac{\xi}{r^4}\bigl(18+e^B+5e^{-B}-8e^{B/2}
-16e^{-B/2}\bigr) +\nonumber \\ \hspace{-2 cm} +\frac{1}{2r^3}\biggl[r^2+12\xi\bigl(1+e^{-B}-2e^{-B/2}\bigr)\biggr]B'
-\frac{4\xi}{r^2}\left(1+e^{-B}-2e^{-B/2}\right)A''+\nonumber \\ \hspace{-2 cm}
+\frac{2\xi}{r^2}\left(1+3e^{-B}-4e^{-B/2}\right)B\, 'A'
-\frac{\xi}{r^2}\left(1+e^{-B}-2e^{-B/2}\right){A'}^2
=0
\label{eq2}
\end{eqnarray}

\begin{eqnarray} \hspace{-2.5 cm}
-2\left[r^2+12\xi\left(1+e^{-B}-2e^{-B/2}\right)\right]A''
+\frac{2}{r}\left[r^2+12\xi\left(1+e^{-B}-2e^{-B/2}\right)\right]B'+\nonumber \\ \hspace{-2.5 cm}
+\frac{4}{r^2}\left[r^2(1-e^B)+4\xi\left(6-e^B+3e^{-B}-8e^{-B/2}\right)\right]
+\biggl[16\xi r\, e^{-B/2}\left(1-e^{-B/2}\right)A''+\nonumber \\ \hspace{-2.5 cm}
+\left\{r^2+12\xi\left(1+3e^{-B}-4e^{-B/2}\right)\right\}B'
+\frac{2}{r}\left\{r^2+4\xi\left(3+7e^{-B}+2e^{B/2}-12e^{-B/2}\right)\right\}\biggr]A'\nonumber \\ \hspace{-2.5 cm}
+\biggl[4\xi r\, e^{-B/2}\left(3e^{-B/2}-2\right)B'
-\left\{r^2+4\xi\left(1-3e^{-B}+2e^{-B/2}\right)\right\}\biggr]{A'}^2+\nonumber \\ \hspace{-2.5 cm}
-4\xi r e^{-B/2}\left(e^{-B/2}-1\right) {A'}^3
=0 \hspace{1.56cm}\label{eq3}
\end{eqnarray}

Notice that the equations obtained are not simple since we have nontrivial coupled differential equations which mix all the functions and their derivatives, namely, $B$, $B'$, $A'$, and $A''$. Some authors have preferred to consider some approximations, restriction of the functions or perturbative calculations even before getting the system of equations. So, equations (\ref{eq1})-(\ref{eq3}) had never been considered explicitly in the literature so far. 

The next step is to obtain an analytical solution which gives us explicitly $P$ and other functions, such as $A'$ or $B'$, so that they can be numerically integrated. After some manipulations and with the help of mathematical software in order to deal with the complex expressions, we were able to reach the following solution:
 
\begin{eqnarray} \hspace{-2 cm}
P=\frac{{\rm e}^{-B}}{8\pi r^4}\Big\{{r}^{2} \left(1-{{\rm e}^{B}} \right) + r^3A'
+12r({{\rm e}^{-1/2\,B}}-1)^2\xi A'+
\nonumber \\ \hspace{-2 cm}
+2{{\rm e}^{B}}({{\rm e}^{-1/2\,B}}-1)^3(3{{\rm e}^{-1/2\,B}}+1)\xi 
+ 2r^2 \xi({{\rm e}^{-1/2\,B}}-1)(3{{\rm e}^{-1/2\,B}}-1) \, {A'}^2\Big\},
\label{press}
\end{eqnarray}

\begin{eqnarray} \hspace{-2 cm}
B'=-\frac{{{\rm e}^{B}}}{r}\bigg\{{r}^{4}\big(1-{{\rm e}^{-B}}-8\,\pi \,{r}^{2}\rho\big)
-96\pi\,{r}^{4}\rho ({{\rm e}^{-1/2\,B}}-1)^2\xi+ 
\nonumber \\    \hspace{-2 cm}
-6\,{r}^{2}({{\rm e}^{-1/2\,B}}-1)^3(5+3{{\rm e}^{-1/2B}})\xi
-8({{\rm e}^{-1/2\,B}}-1)^5(11+9{{\rm e}^{-1/2B}})\xi^2 +
\nonumber \\    \hspace{-2 cm}
-8\,{r}^{3}{{\rm e}^{-1/2\,B}}({{\rm e}^{-1/2\,B}}-1)\Bigl[8\,\pi \,{r}^{2}\rho
+({{\rm e}^{-1/2\,B}}-1)(2{{\rm e}^{-1/2\,B}}+1)\Bigr]\,A'\xi+
\nonumber \\    \hspace{-2 cm}
-16\, r\, {{\rm e}^{-1/2\,B}}({{\rm e}^{-1/2\,B}}-1)^4(9{{\rm e}^{-1/2\,B}}+5)\,A'\xi^2
+2\,r^4 {{\rm e}^{-B}}({{\rm e}^{-1/2\,B}}-1)^2  \,{A'}^2\xi+
\nonumber \\    \hspace{-2 cm}
 -8\,r^2 {{\rm e}^{-B}}({{\rm e}^{-1/2\,B}}-1)^3 (9{{\rm e}^{-1/2\,B}}-1)\,{A'}^2\, \xi^2\bigg\}
\bigg/ 
 \bigg\{ r^4+16\,{r}^{2}({{\rm e}^{-1/2\,B}}-1)^2\xi+
\nonumber \\   \hspace{-2 cm}
+48\,({{\rm e}^{-1/2\,B}}-1)^4\xi^2
+16\,{r}^{3}{{\rm e}^{-1/2\,B}}({{\rm e}^{-1/2\,B}}-1) A'\xi +
\nonumber \\   \hspace{-2 cm}
+  96\,r\,{{\rm e}^{-1/2\,B}}({{\rm e}^{-1/2\,B}}-1)^3 A'\xi^2 
   +48\,{r}^{2}{{\rm e}^{-B}}({{\rm e}^{-1/2\,B}}-1)^2
 {A'}^2\xi^2 \bigg\}, 
\label{BL}
\end{eqnarray}
and
\begin{eqnarray} \hspace{-2.5 cm}
A''={\rm e}^B\bigg\{{r}^{4} \left( 8\,\pi \,{r}^{2}\rho+3\,{{\rm e}^{-B}}-3 \right) 
 -2r^2\big(96\,{{\rm e}^{-1/2\,B}}\pi \,{r}^{2}\rho 
 -48\,{{\rm e}^{-B}}\pi \,{r}^{
2}\rho-48\,\pi \,{r}^{2}\rho +\nonumber \\ \hspace{-2.5 cm}  -44\,{{\rm e}^{-1/2\,B}}-25\,{{\rm e}^{-2
\,B}}
+52\,{{\rm e}^{-3/2\,B}}-6\,{{\rm e}^{-B}}+23
\big) \xi+24\big(-5 -30\,{{\rm e}^{-5/2\,B}}+\nonumber \\ \hspace{-2 cm}
+18{{\rm e}^{-1/2\,B}} +45{{\rm e}^{-2B}}
+7{{\rm e}^{-3B}}-15{{\rm e}^{-B}}-20{{\rm e}^{-3/2B}}
\big)\xi^2 
+\bigg[\dfrac{1}{2}\,{r}^{5} \left( 8\,\pi \,{r}^{2}\rho+3\,{{\rm e}^{-B}}-1 \right) \nonumber \\ \hspace{-2.5 cm} -r^3\big(192\,{{\rm e}^{-1/2\,B}}\pi \,{r}^{2}\rho 
-144\,{{\rm e}^{-B}}\pi \,{r}
^{2}\rho-48\,\pi \,{r}^{2}\rho-72\,{{\rm e}^{-1/2\,B}}+22\,{{\rm e}^{-
B}}+\nonumber \\ \hspace{-2.5 cm}
-77\,{{\rm e}^{-2\,B}}+112\,{{\rm e}^{-3/2\,B}}+15
  \big)\xi+4r \big( 111\,{{\rm e}^{-3\,B}}
 -396\,{{\rm e}^{-5/2\,B}} +92\,{{\rm e}^{-1/2\,B}
} +\nonumber \\ \hspace{-2.5 cm} -147\,{{\rm e}^{-B}}+463\,{{\rm e}^{-2\,B}}
-112\,{{\rm e}^{-3/2\,B}}-11
 \big)\xi^2
 \Big]A'+\biggl[-\frac{1}{2}{{\rm e}^{-B}}{r}^{6} -4r^4\big( 2\,{{\rm e}^{-B}}+\nonumber \\ \hspace{-2.5 cm}
 +8\,{{\rm e}^{-1/2\,B}}\pi \,{r}^{2}\rho -12\,{{\rm e}^{-B}}\pi \,{r}^{2
}\rho-{{\rm e}^{-1/2\,B}}-5\,{{\rm e}^{-2\,B}}
+4\,{{\rm e}^{-3/2\,B}} \big) \xi +\nonumber \\ \hspace{-2.5 cm} +4r^2 \big(10\,{{\rm e}^{-1/2\,B}}+242\,{{\rm e}^{-2\,B}}+93\,{{\rm e}^{-3\,B}}
-270\,{{\rm e}^{-5/2\,B}}-36\,{{\rm e}^{-3/2\,B}}-39\,{{\rm e}^{-B}}
\big)\xi^2\Bigr]{A'}^2+\nonumber \\ \hspace{-2.5 cm}
 +{\rm e}^{-B}\xi r^3\Bigl[ r^2\big(8\,{{\rm e}^{-B/2}}-7\,{{\rm e}^{-B}}+1 \big) +4 \big(34\,{{\rm e}^{-B}}
 +21\,{{\rm e}^{-2\,B}}-48\,{{\rm e}^{-3B/2}}-8{{\rm e}^{-B/2}}+1 \big)\xi\Bigr]{A'}^3\nonumber \\ \hspace{-2.5 cm}
  -12r^4\big( {{\rm e}^{-3\,B}}+{{\rm e}^{-2\,B}}-2\,{{\rm e}^{-5/2\,B}}\big)\xi^2{A'}^4\bigg\}\bigg/ 
  r^2\bigg\{ r^4+16\big({r}^{2}-2\,{{\rm e}^{-1/2\,B}}{r}^{2}+{{\rm e}^{-B}}{r}^{2} \big)\xi+\nonumber \\ \hspace{-2.5 cm}
  +48\big(1-4{{\rm e}^{-1/2B}}+6{{\rm e}^{-B}}-4{{\rm e}^{-3/2B}
} 
+{{\rm e}^{-2B}}\big)\xi^2+\Big[16r^3\big( {{\rm e}^{-B}} -{{\rm e}^{-1/2\,B}}\big)\xi +\nonumber \\ \hspace{-2.5 cm}
-96r  \big( {{\rm e}^{-1/2\,B}}-3\,{{\rm e}^{-B}}+3\,{{\rm e}^{-3/2\,B}}-\,{
{\rm e}^{-2\,B}}  \big) \xi^2\Big]A'
+48r^2\big( {{\rm e}^{-B}}-2\,{{\rm e}^{-3/2\,B}}+{{\rm e}^{-2\,B}}\big)\xi^2 {A'}^2 \bigg\}. \nonumber \\ 
\label{A2L}
\end{eqnarray}
Notice that, although we have obtained the expression just above for $A''$ in terms of the independent variables, one does not necessitate to use it, since it will be more useful to consider alternatively the conservation equation (\ref{ce1}) which relates $A'$, $P'$ and $\rho$ instead, since the latter is a much simpler expression.

The energy density $\rho$ is related to the ``mass interior to radius r", $m(r)$, as follows:
\begin{eqnarray}
\frac{dm}{dr}=4\pi \rho \, r^2.
\label{dmdr}
\end{eqnarray}

{It is worth mentioning that equation (\ref{dmdr}) is the same that appears in the derivation of TOV GR, which is also adopted explicitly or implicitly by other authors  (see, e.g., Ref. \cite{Ilijic20}). Although, it is worthy of note that some authors consider that the calculation of mass in $f(T)$ gravity is still an open problem (see, e.g., Refs. \cite{Ilijic} and \cite{olmo}).}

{Having in mind that it is not an observable, the total rest mass $M_0$ is an interesting quantity to calculate, since it can be used to make comparisons between different theories. For example, for a given central density and a given equation of state, different theories must provide different values of $M_0$.

Recall that the total rest mass $M_0$ is obtained by solving 
\begin{eqnarray}
\frac{dm_0}{dr}=4\pi \rho_0 \,e^{B/2} r^2 \, ,
\label{dm0dr}
\end{eqnarray}
where $\rho_0$ is the rest mass density and $4\pi e^{B/2} r^2 dr$ is the proper volume element.
}

{ 
Note that the calculation of $M_0$ in $f(T)$ gravity is unambiguously given, since it only takes into account the total rest mass.  The internal energy and the gravitational potential energy do not enter in the calculation of $M_0$. On the other hand, in the calculation of $M$ these forms of energy referred to above are taken into account. Since gravity is now modified, this is why there is no guarantee that $M$ accounts for the mass of a compact object in $f(T)$ gravity. From now on, M will be called $M_{GR}$, as this mass calculation is as done in GR.}

{
An interesting  way to calculate the mass is related to the obvious fact that the mass of the object appears in some way in the metric it generates. That is, the mass is encoded in $A(r)$ and $B(r)$. Consequently, one can find the mass of an asymptotically Minkowskian metric, i.e., the ``mass measured by an observer at infinity'' ($M_{I\! N\! F}$). Obviously, applying this procedure in GR one obtains $M$, since the mass that appears in the Schwarzschild metric is nothing but the``mass measured by an observer at infinity''.}

{ 
To begin with, notice that equation (\ref{BL}) also holds for vacuum, since $\rho$ smoothly goes to zero. Now, considering equation (\ref{BL}) for $r \rightarrow \infty$, one obtains
\begin{equation}
    B' = - \frac{B}{r} \qquad {\rm for \; r \rightarrow \infty},
\end{equation}
whose solution reads
\begin{equation}
    B(r) = \frac{C}{r}, \label{BA}
\end{equation}
where $C$ is a positive constant.
}

{
Equation (\ref{press}) also holds for vacuum, since $P$ smoothly goes to zero just like $\rho$. Now, considering equation (\ref{press}) for $r \rightarrow \infty$ and using (\ref{BA}), one obtains
\begin{equation}
    A' = \frac{B}{r} = \frac{C}{r^2} \qquad {\rm for \; r \rightarrow \infty},
\end{equation}
whose solution reads
\begin{equation}
    A(r) = - \frac{C}{r}, \label{AA}
\end{equation}
where we consider that $A(r \rightarrow \infty ) = 0$ to set the other integration constant equal to zero. Notice that no terms in $\xi$ appear in the asymptotic solutions for $A(r)$ and $B(r)$.
}

{
To proceed, we write the metric for r $\rightarrow \infty$, namely 
\begin{equation}
    ds^2=(1+A(r))\, dt^2-(1+B(r))\, dr^2-r^2\, d\theta^2-r^2 \sin ^2 \theta \, d\phi^2 
\end{equation}
that can be rewritten, with the use of equations (\ref{BA}) and (\ref{AA}), as follows
\begin{equation}
    ds^2=\left(1- \frac{C}{r}\right)\, dt^2-\left(1 + \frac{C}{r}\right)\, dr^2-r^2\, d\theta^2-r^2 \sin ^2 \theta \, d\phi^2 \label{metric},
\end{equation}
which strongly suggests that $C$ is related to the ``mass measured by an observer at infinity'' ($M_{I\! N\! F} $). Then, we set $C \equiv 2 M_{I\! N\! F} $. This is just like Schwarzschild metric very far from the source.
}

{
In practice, to obtain $M_{I\! N\! F}$, one solves equations (\ref{press}) and (\ref{BL}) for vacuum, starting from the star's surface till, say, hundred or thousand times the star's radius. This ensures that equation (\ref{BA}) holds.
Thus, $M_{I\! N\! F} = C/2$ is easily obtained, namely,
\begin{equation}
    M_{I\! N\! F} = \frac{C}{2} =  \lim_{r\to\infty} \frac{1}{2}\, r\,B(r).
    \label{minf}
\end{equation}
}

{
It is worth mentioning that in the models presented in Section \ref{sec 4}, both $M_{I\! N\! F}$ and $M_{GR}$ are calculated. It will be seen that depending on the value of $\xi$, $M_{I\! N\! F}$ and $M_{GR}$ are significantly different.
}

Before closing this section, note that for $f(T)=T$, one should recover the TOV {equation that comes} from General Relativity (see \ref{app} for detail). 

{In summary, to model stars on the particular $f(T)$ gravity adopted here for a given equation of state (EOS), one needs to integrate the following system of differential equations: (\ref{ce1}), (\ref{press}), (\ref{BL}) and (\ref{dmdr}).}

As we are interested in modelling neutron stars, we will need to consider their EOS, which can be quite complex. For this reason, it is common to start by considering the so-called polytropic equations instead of the realistic ones. In general, the polytropic offers a very consistent description of these objects \cite{eos0}. For more information and examples of neutron star state equations, see \cite{eos1}-\cite{eos3}.

\section{Numerical examples} 
\label{sec 4}

\subsection{Polytropics}
\label{sec 4.1}
The main proposal here is to show that, despite the apparent complexity of the equations obtained in last section, we can apply them to cases of physical interest.

As an practical example, we consider an specific polytropic equation of state \cite{eos0}, since  it  can  offer  a  simpler  and satisfactory description for neutron stars, for example.

Recall that the polytropic reads
\begin{equation}
    P = k\, {\rho_o}^ \gamma,
    \label{tpol}
\end{equation}
where $\rho_o$ is the rest-mass density, $k$ is the polytropic gas constant and $n$ defined by $\gamma \equiv 1+1/n$ is the polytropic index. From the first law of thermodynamics one obtains the mass-energy density, $\rho$, namely $\rho = \rho_o + n\,P$ (see, e.g., \cite{BS2010}).

{Since our aim here is to compare how different $f(T)$ is in comparison to General Relativity regarding compact objects, we can proceed by solving equations (\ref{ce1}), (\ref{press}), (\ref{BL}) and (\ref{dmdr}) to model a compact star for the $n=1$ polytropic equation of state, $P={\rho_0}^2$, where we have set all quantities in nondimensional form\footnote{{In \cite{BS2010} it is shown that, in geometrized units, $k^{n/2}$ has units of length. Therefore, one can define nondimensional quantities, namely, $\bar{r} = k^{-n/2}r $, $\bar{P} = k^{n}P $, $\bar{\rho} = k^{n}\rho $, $\bar{M} = k^{-n/2}M $ and $\bar{T} = k^{n}T $. To simplify the notation, we then omit the bars in our equations. }}. In practice, this is equivalent to set $k = G = c = 1$. }

In order to perform the numerical integration, we have written a numerical code in Python. The procedure is essentially to integrate numerically the differential equations for $m(r)$ and $P(r)$ for a given EOS. Then, it is necessary to choose a central density, $\rho_c$, and using the EOS one obtains the central pressure $P_c$. In addition, providing the central boundary conditions
\begin{equation}
    m = 0  \quad {\rm and}  \quad P = P_c \quad {\rm at} \quad r =0,
\end{equation}
one obtains $m(r)$, $P(r)$ and $\rho(r)$, i.e., the structure of the star. The radius of the star, $R$, is given by the value of r for which $P(r) = 0$. That is, one starts the integration of the set of differential equations at $r=0$ and continues it till the value of $r$ for which $P(r)=0$. Finally, the mass of the star {calculated just like in GR } is then given by $M_{GR} = m(R)$. {The calculation of $M_{I\! N\! F}$ is given by equation (\ref{minf}), and to do so equations (\ref{press}) and (\ref{BL}) are solved for vacuum, starting from the star's surface till, say, hundred or thousand times the star's radius.}

{Here, regarding to the modelling of a compact star, there is an additional equation in comparison to General Relativity, namely, the differential equation for $B(r)$.} Then, one needs to provide a central boundary condition for $B$. {Hence, regularity at the center implies that $B_c = 0$.}

\begin{figure*}
    \includegraphics[scale=0.32]{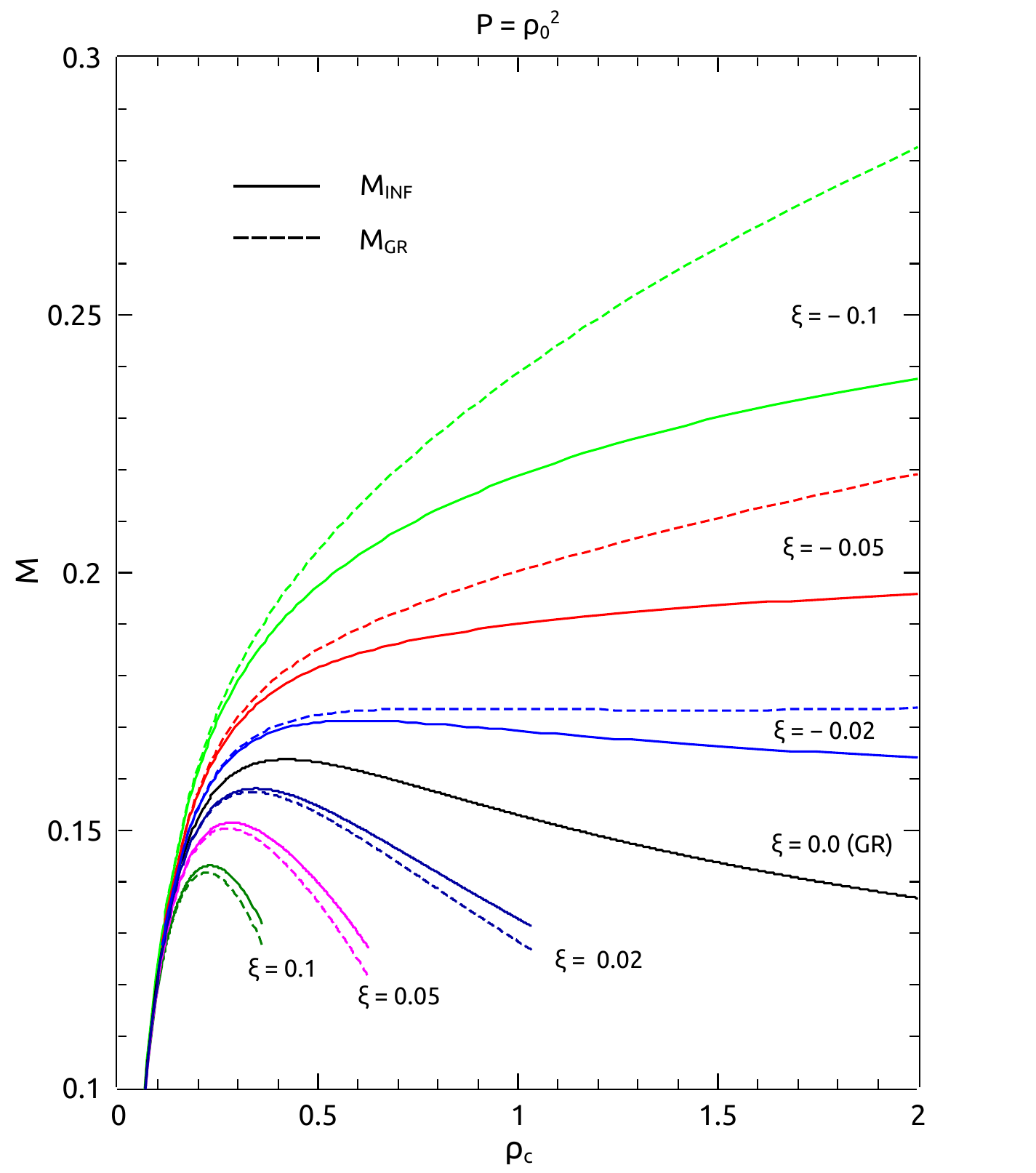}
    \includegraphics[scale=0.32]{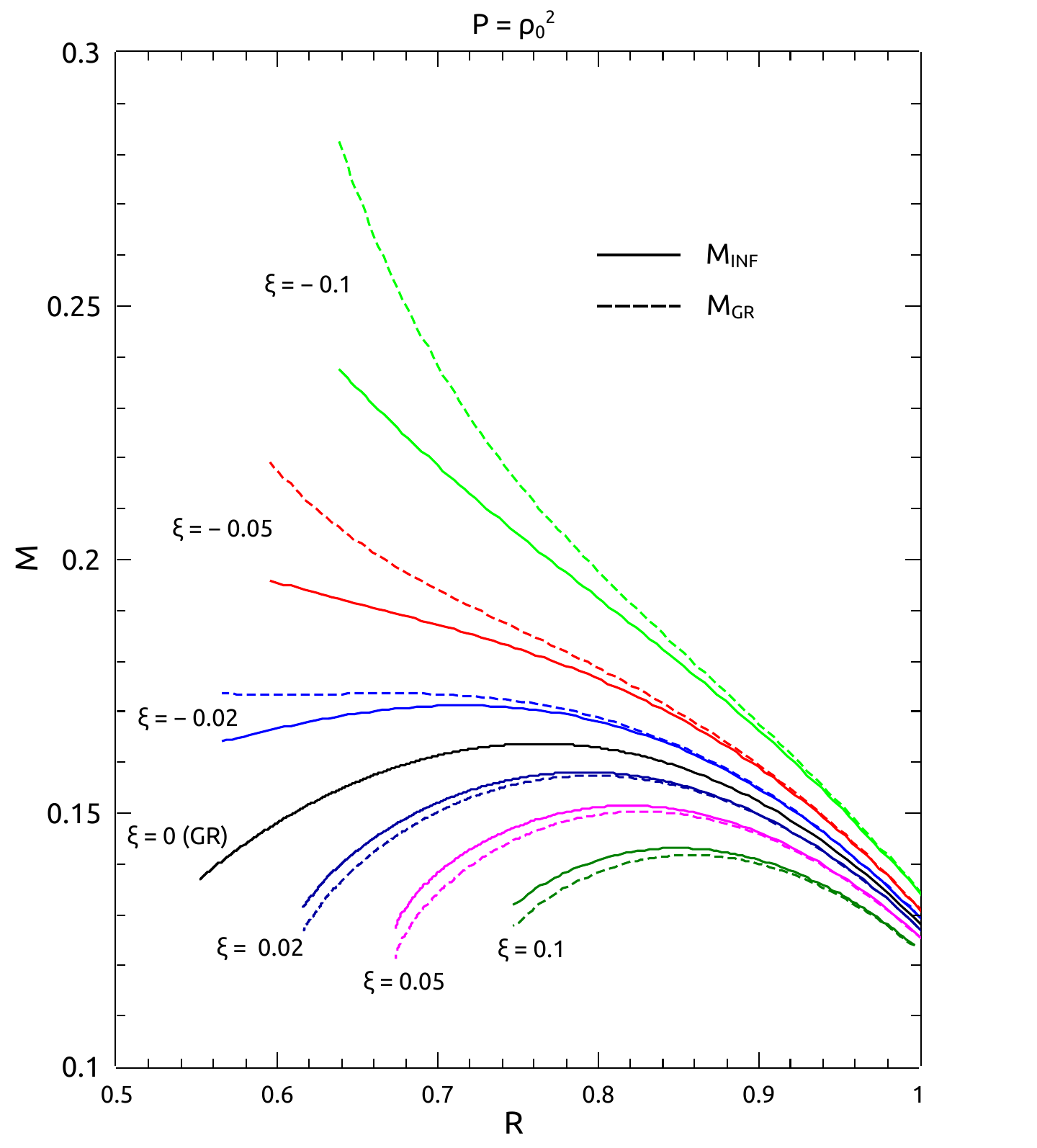}
\caption{Right (left): sequences of M$_{RG}$ and M$_{I\! N\! F}$ vs. central mass-energy \\ density $\rho_c$ (radius R) for $P= \rho_0^2$ and different values of $\xi$.}
    \label{PPN1MD}
\end{figure*}

{As a result of this numerical integration, different models of gravity from $f(T) = T + \xi T^2$ (with different values for $\xi$) could be compared. In figure \ref{PPN1MD}, we see that, for a range of values of $\rho_c$, one obtains the corresponding masses and radii and, consequently, the ``Mass $\times$ Radius'' and ``Mass $\times$ $\rho_c$'' curves. From these curves, we can identify, for example, the maximum mass allowed for a given EOS. For $\xi=0$, we have $f(T)=T$, which is nothing but the TEGR, whose TOV equations after a simple manipulation are reduced to TOV GR.}

{Looking at Figure \ref{PPN1MD} more closely, we can notice that for $\xi \ge 0$ there are maximum masses {(for both $M_{GR}$ and $M_{I\! N\! F}$) }. In addition, the greater $\xi$ is, the lower the maximum masses. Considering now that $\xi$ could have negative values, one clearly see that, as compared to GR, the masses for a given $\rho_c$ increases for decreasing values of $\xi$.}

{Additionally, one can see from figure \ref{PPN1MD} how the way of calculating the total mass affects the behavior mentioned above. The dashed (continuous) lines represent the curves obtained using $M_{GR}$ ($M_{I\! N\! F}$). For $\xi \ge 0$, we have greater maximum masses if the mass is given by $M_{I\! N\! F}$. Also, for $\xi<0$, the greater the module of $\xi$, the more difference between the curves for $M_{RG}$ and $M_{I\! N\! F}$ when $\rho_c$ increases.}

{Notice that this very polytropic was studied by \cite{Ilijic} and a detailed comparison  shows that the results are in perfect agreement. {However, in \cite{Ilijic} the authors mentioned that numerical instabilities appear for models with $P_c/\rho_c$ above given values. These instabilities also appear in our calculations but only for $\xi \ge 0$.  The instabilities reported in \cite{Ilijic} for $\xi < 0$ do not appear in our calculations.}
It is noteworthy that these authors adopted the covariant formulation and wrote the set of differential equations to be solved differently from what we did here, so our approaches are not the same, although they are equivalent.
{A very important difference between our approaches has to do with the {\it conservation equation}, since they do not consider it. It seems that they did not realised that it holds in $f(T)$ gravity, or they did not pay attention to this issue.} }

The use of the {\it conservation equation} simplifies the numerical calculations\footnote{{The expression for $A'$ can be easily obtained from the equation (\ref{press}), eliminating the need for a numerical integration. This fact makes all the calculation simpler and more stable than those presented in \cite{Ilijic}.}}, and this is probably the reason why instabilities do not appear in our approach for $\xi < 0$. Consequently, we can model stars for any central density for $\xi < 0$. 
Notice that, in left panel of figure \ref{PPN1MD}, the calculations are presented up to $\rho_c = 2$ (i.e.,  $\sigma_0 = P_c/\rho_c = 0.5$ using the definition of Ref. \cite{Ilijic}) for convenience, but we can model stars for larger values than this central density without any numerical instability. {Thus, a question that could be asked on the basis of Ref. \cite{Ilijic} can be adequately answered, namely, whether or not there are indeed maximum masses for $\xi < 0$. The novel approach presented here shows that in fact there is no maximum mass for {some of} the negative values of $\xi$ that we studied.}
{Therefore, our approach has allowed us to go one step further.}

{Furthermore, in contrast to previous works in the literature, we have explicitly showed the system of equations to be solved, which is commonly avoided by other authors due to its complexity. Here, we have not only presented the complete system of equations which can be useful for a future application of this method to more general forms of $f(T)$ models, but also we did so without any approximations or restrictions for simplicity usually made until now.}

\begin{figure*}
\includegraphics[scale=0.32]{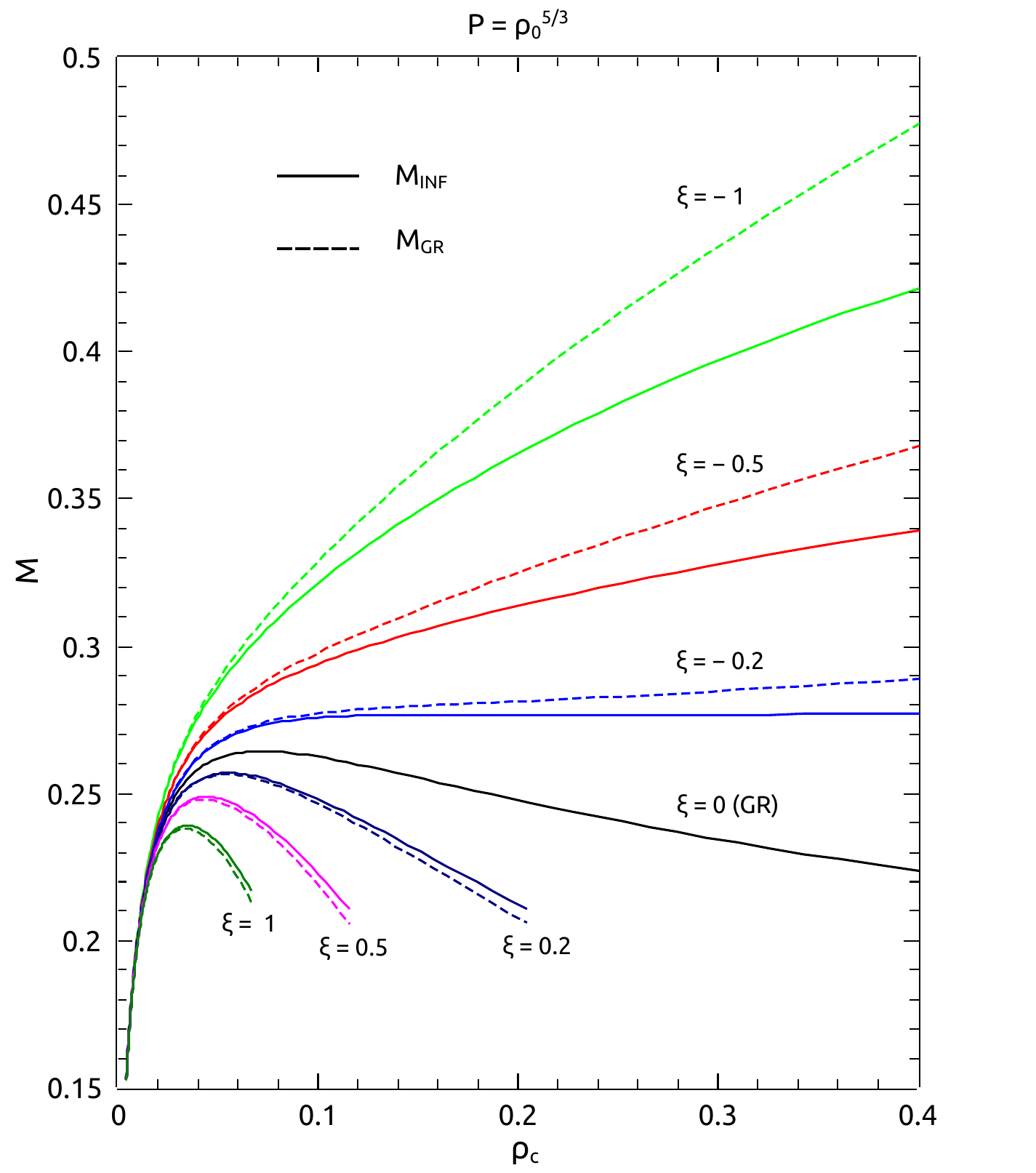}
\includegraphics[scale=0.32]{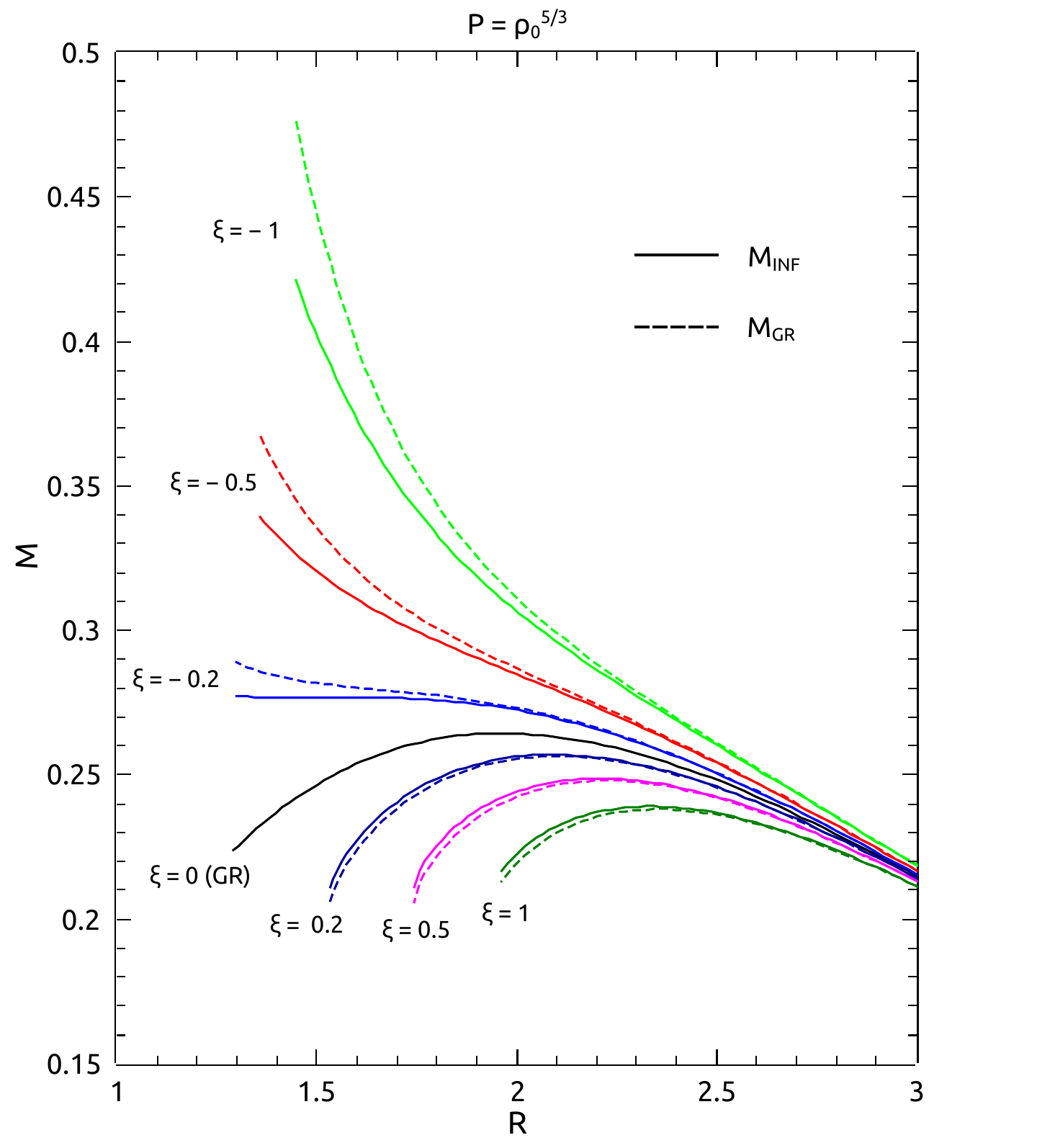}   
\caption{The same as figure \ref{PPN1MD} now for $P= \rho_0^{5/3}$.}
\label{PPN5s3MD}
\end{figure*}

In figure \ref{PPN5s3MD} are shown  ``Mass $\times$ Radius'' and ``Mass $\times$ $\rho_c$'' sequences, respectively, now for a polytropic with n = 3/2, i.e., $P={\rho_0}^{5/3}$ (non-relativistic Fermi gas). As can be seen, the behaviour is qualitatively the same as those for n = 1 ($P={\rho_0}^2$) polytropic.

{The numerical integration of equations (\ref{ce1}), (\ref{press}), (\ref{BL}) and (\ref{dmdr}) was entirely performed using packages from Python programming, more specifically, we have used  `solve-ivp' (see, e.g., \url{https://docs.scipy.org/doc/scipy/reference/generated/scipy.integrate.solve_ivp.html##scipy-integrate-solve-ivp} for details) with integration method option ‘RK45’ (explicit Runge-Kutta method of order 5(4)). However, all other possible integration options of  `solve-ivp' were tested, namely, `OP853', `RK23', `Radau', `BDF' and `LSODA'.}

It is worth mentioning that we consider  TOV in $f(T) = T + \xi T^2$ in two other articles. We refer the reader to these papers for further models in this particular $f(T)$. In one of them we study other polytropics \cite{FA2}. In the other one, we study some realistic equations of state in modeling neutron stars \cite{FA3}.

\subsection{Torsion}
\label{torsion}

{As is well known, in the Teleparallel theory of Gravity, which is equivalent to GR, the torsion scalar $T$ is considered instead of the Ricci scalar in its formulation. An obvious fact in GR is that the Ricci scalar outside (vacuum) of any matter distribution is zero. One would think of that the torsion would be also null. However, having a look, for example, at equation (\ref{008}), one can easily conclude that $T$ is not null outside a spherically symmetrical distribution.}

{On the other hand, at the center of a star, the Ricci scalar is not zero and, instead, the torsion scalar is.} {Using equations (\ref{press}) and (\ref{BL}) for $r \rightarrow 0$ in equation (\ref{008}), one obtains that 
$T(r) \propto -r^2$ for $r \rightarrow 0$. Thus, one easily concluded that $T(0)=0$. It is worth noting that this dependence of $T(r)$ for small $r$ is corroborated by our numerical calculations.}

{As far as we know, the papers related to compact objects in $f(T)$ gravity do not present explicitly the behaviour of torsion inside and outside the matter distribution. Here, we present $T(r)$ for a couple of models. This will be useful to show, compared to GR, how much the gravitational interaction is more or less intense for positive and negative values of $\xi$. One could think of that torsion is closely related to the gravitational interaction.}

\begin{figure}
\centering
\includegraphics[scale=0.4]{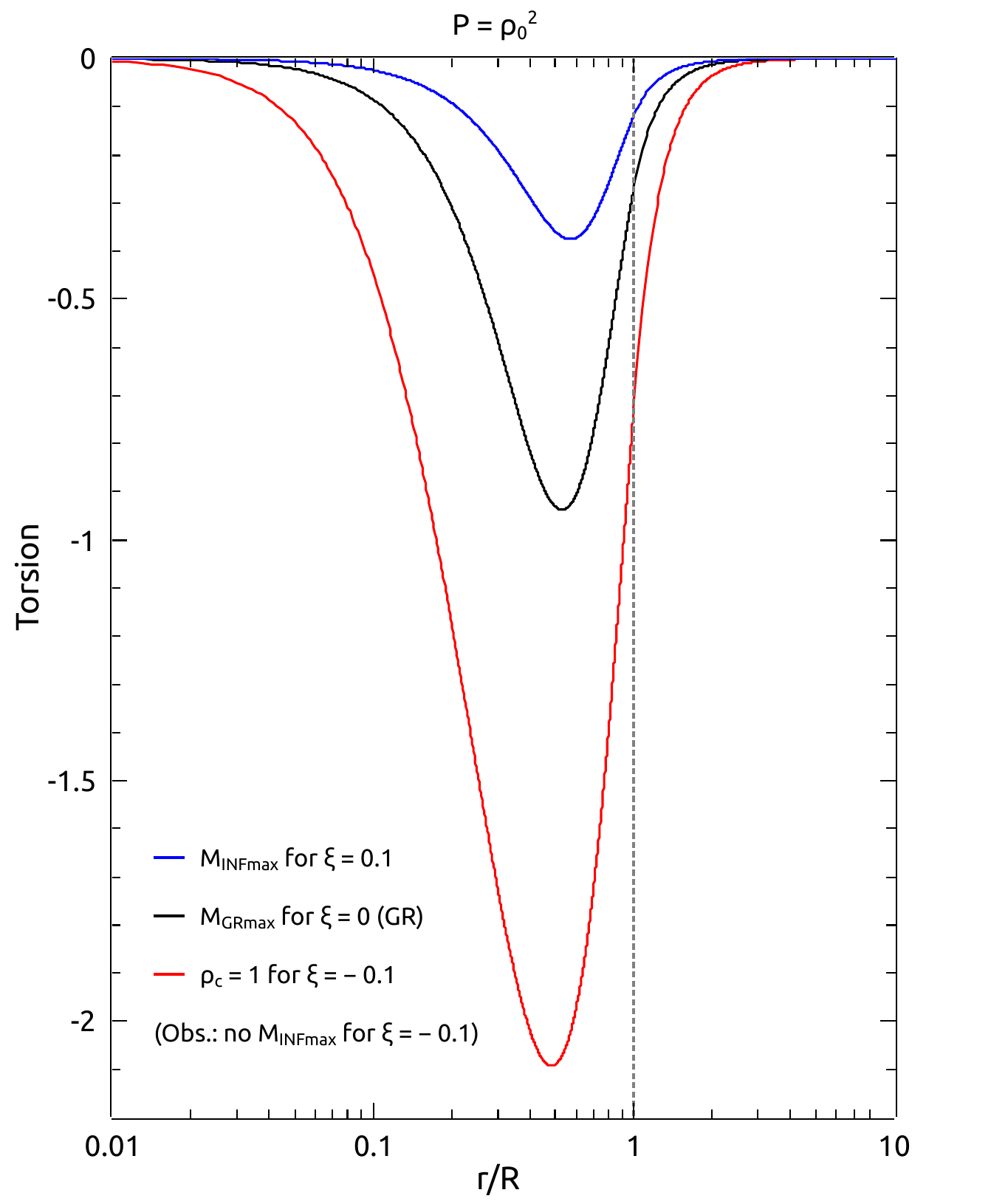}
\caption{Torsion as a function of r inside and outside the {polytropic} $P= \rho_0^2$ }
    \label{Torsion}
\end{figure}

{In figure \ref{Torsion}, it is shown the (dimensionless) torsion as a function of the $r$ coordinate  inside and outside a spherically symmetrical distribution of polytropic EOS $P= \rho_0^2$ for $\xi = - 0.1$, 0 and 0.1.}

{Notice that the torsion is negative inside and outside the matter distribution considered and has a shape similar to a ``potential well". Moreover, the numerical calculations show that the torsion is null at $r = 0$. }

{For $\xi = 0$ and 0.1, we consider the behaviour of the torsion for the maximum masses, for which the torsion is more pronounced. For masses smaller than the maximum masses, the torsion curves would be above the blue ($\xi = 0.1$) and black ($\xi = 0$) curves (see figure \ref{Torsion}), as expected. The ``potential wells" are shallower for masses below the maximum masses.}

{Since for $\xi = - 0.1$ there is no maximum masses, we consider the torsion for $\rho_c$ =~ 1. For central densities $\rho_c < 1$ ($\rho_c > 1$), the torsion curves are above (below) the red line. 
The ``potential wells" are shallower (deeper) for $\rho_c < 1$ ($\rho_c > 1$).}

{One could think of that the gravitational interaction is more intense for decreasing values of $\xi$. This can be understood via equation \ref{fT}. The more negative $\xi$ is, the more negative $f(T)$ is and, therefore, the more intense the gravitational interaction.} {On the other hand, the gravitational interaction is less intense for positive values of $\xi$.}

{It is worth mentioning that the equations used to model stars presented here are also valid outside the matter distribution since P and $\rho$ go smoothly to zero. Therefore, there is no jumps in the potentials $A$ and $B$ at least in their first derivatives. That is why $T(r)$ and its first derivative have no jumps at $r = R$.}

{Another interesting fact is the $T(r)$ goes to zero outside the star at a distance of only a few radii from the star. For example, for $\xi=-0.1$ and $\rho_c$ = 1, we obtain  $T(3R) \sim T(R)/100$. Thus, the spacetime becomes almost flat relatively near the star surface as in GR.}

\begin{figure}
\centering
\includegraphics[scale=0.4]{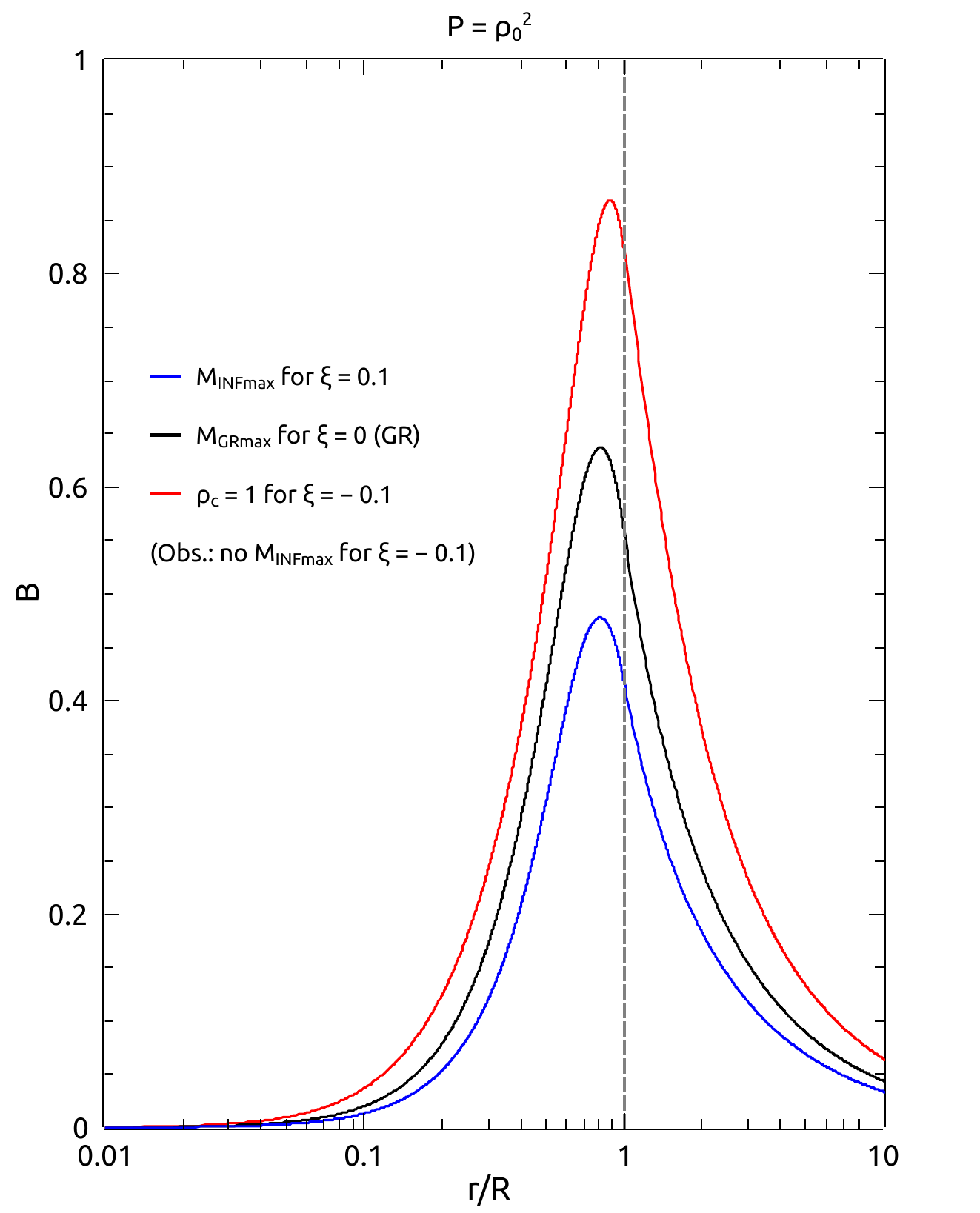}
\caption{$B(r)$ inside and outside the polytropic $P= \rho_0^2$ }
    \label{Br}
\end{figure}
{Before closing this section, we show in figure \ref{Br}, as an example, the metric function $B(r)$, which is one of the outputs of our calculations,  inside and outside of a spherically symmetrical distribution of polytropic EOS $P= \rho_0^2$ for $\xi = - 0.1$, 0 and 0.1. Note that, the curve for $\xi = - 0.1$ (0.1) is above (below) the GR curve. This is because the gravity is more intense (weaker) than that of GR. Outside the polytropic, $B(r)$ goes to zero for distances of a few radii from the stars. This in an indication that the spacetime becomes flat not so far from the star surface.} 

\section{Final Remarks}\label{sec 5}

{Modelling spherical stars has been a much-addressed topic in the recent literature. At the same time, alternative descriptions of gravity have been highly pursued due the open problems in this area. In view of that, it is always productive offer different approaches do deal with the problem and even more useful if the procedure is quite general and able to be reproduced for other proposals. The novel approach presented here, despite the considerable equations that appear, can be exactly repeated for other models. Additionally, one of our concerns was to explicit each step of the method in order to facilitate future works with the same approach, in contrast of some authors who have adopted effective strategies but do not detail the procedure sufficiently.}

As we showed in Section \ref{sec 3}, for the particular case in which $f(T) = T + \xi T^2$, the equations to be solved are very complicated. Nevertheless, the correspondent TOV equations could be obtained without any restrictions. It is worth stressing that we adopted here a particular form for $f(T)$ as an example, but since the procedure considered here is quite general, any $f(T)$ can be in principle considered. Obviously, one is also free to choose the equation of state. 

For the present choice of $f(T)$, one can then solve equations (\ref{ce1}), (\ref{press}), (\ref{BL}) and  (\ref{dmdr}) to model a compact star for a given equation of state. Notice that in the way these equations are obtained, one does do not need to assume beforehand any particular form for $A$, $B$ and $T$, as some papers in the literature do.

One could ask oneself why this particular set of equations has been chosen. Notice also that in the present context,  by setting $\xi =0$ into equations (\ref{press}) and (\ref{BL}), one obtains {(see \ref{app})} (\ref{ARG}) and (\ref{BRG}), respectively. Therefore, one could think of equations (\ref{press}) and (\ref{BL}) as a kind of generalisation of (\ref{ARG}) and (\ref{BRG}). It is worth mentioning that, if we use (\ref{ce1}), equation (\ref{A2L}) is no longer necessary since it does not contain any new information. Moreover, it can well be obtained by a combination of the other equations that in fact will be resolved.
 
Another interesting issue concerning the equations is that the combination of equations (\ref{ce1}) and (\ref{press}) is a generalisation of equation (\ref{HI}). It was precisely this fact that has allowed the resolution of the system in the current approach without any further complications.

{Regarding to maximum masses, we have obtained that for $\xi>0$ the greater $\xi$ is, the lower the maximum mass. For the negative values of $\xi$ that we studied, there is no maximum mass. This question came up in \cite{Ilijic} but it has not been answered because there was instabilities for $\xi<0$. In our approach, this instability is absent. Possibly, this is due the use of conservation equation instead of the differential equation for $A'$ which come from the TOV equations. Consequently, we can model stars for any central density for $\xi < 0$. {Moreover, the way of calculating the total mass affects slightly the ``Mass $\times$ Radius'' and ``Mass $\times$ $\rho_c$'' curves and it seems to be more physically reasonable to consider the mass given by $M_{I\! N\! F}$ since it is the mass calculated by an observer at the infinity, without all the possible local effects.}} 

Also, it is worth noticing that $e^{-B(r)} = 1 - 2\,m(r)/r$ in TOV GR, whereas here one has to numerically integrate equation (\ref{BL}) to obtain $B(r)$.
This integration can be performed
without any kind of approximation.

Once the equation of state has been chosen, the first numerical test to be performed would be with $ \xi = 0 $, since, for this value, we should retrieve results already known from the RG TOV equations, for example, the maximum mass allowed for a neutron star for the chosen equation of state. Then, different values of $\xi$ should be taken, leading to different $f(T)$ gravity models applied to neutron stars.

{Then, using the packages from Python programming, more specifically, `solve-ivp' (see, e.g., \url{https://docs.scipy.org/doc/scipy/reference/generated/scipy.integrate.solve_ivp.html##scipy-integrate-solve-ivp} for details) with integration method option ‘RK45’ (explicit Runge-Kutta method of order 5(4)),} it is possible implement a program that, starting from a given central density for the neutron star and the specific equation of state, could lead us to the structure of the star, i.e., $ P $, $\rho$ and $m$ as a function of $r$, as was done in the previous section for the $n=1$ and 3/2 polytropics. This allows us, among other things, to obtain the value of $ r $ for which the pressure $ P $ vanishes, which would correspond to the external limit of the star, i.e., its radius $R$ and, consequently, the mass $M = m(R)$ is also obtained. Repeating the procedure for different central densities, it is possible to obtain a curve of the type `$`Mass \; \mbox {vs.} \; Radius $'', which would indicate the maximum mass allowed by the model for a given equation of state.

Due to the complexity of the differential equations, all this process of finding the best integrator and performing the numerical analysis is quite elaborate, but it has been shown perfectly possible to obtain its numerical solution, as we have seen from the examples considered previously. Analysis for more general equations of state will be presented in future works to appear elsewhere, either for polytropic and realistic equations of state.

{Additionally, we have discussed the behaviour of torsion inside and outside the matter distribution for different models. This can tell us about the intensity of the gravitational interaction for different $\xi$ and some results could be compared to those from GR.}

\appendix
\section{Recovering TOV GR}\label{app}

Although this discussion appears in related studies, it is worth showing explicitly that one can recover TOV GR by setting $f(T) =T$. As is well known, such a particular case is nothing but the teleparallel equivalent of GR.

By substituting  $f(T) = T$ into equations (\ref{026}) and (\ref{025}), we obtain

\begin{equation}
    8\pi p r^2 = - 1 + e^{-B} + r\,A'\,e^{-B}
\label{ARG}    
\end{equation}
and
\begin{equation}
    8\pi\rho r^2 = 1 - e^{-B} + r\,B'\,e^{-B}
\label{BRG}
\end{equation}

Integrating equation (\ref{BRG}) one obtains

\begin{equation}
    e^{-B} = 1 - \frac{2m}{r},
\label{B}    
\end{equation}
\noindent where 
\begin{equation}
\frac{dm}{dr} = 4\pi\rho r^2.
\label{mr}
\end{equation}

Using these last two equations in (\ref{ARG}), one obtains 
\begin{equation}
A' = 2 \,\frac{4\pi pr^3+m}{r^2- 2\,m\,r}
\end{equation} 

Since from the {\it Conservation Equation} (see, e.g., Ref. \cite{Bohmer}) one has
\begin{equation}
    A' = -2 \frac{p'}{p+\rho},
\label{ce}    
\end{equation}
the equation for the hydrostatic equilibrium reads
\begin{equation}
p' = - (p+\rho)\,\frac{4\pi pr^3+m}{r^2- 2\,m\,r}.
\label{HI}    
\end{equation}

Equations (\ref{B}) - (\ref{HI}), as expected, are the equations one has in general relativity, namely, TOV GR.

\ack
J.C.N.A. thanks FAPESP (2013/26258-4) and CNPq (308367/2019-7) for partial financial support. The authors would like to thank Rafael da Costa Nunes for discussions related to the $f(T)$ theory. {Last but not least, we thank the referee for the detailed and valuable review of our article, which helped to substantially improve it.}

\end{document}